\documentstyle[prb,aps,psfig,multicol]{revtex}

\newcommand{\crossprod}{\times}
\newcommand{\be}{\begin{equation}}
\newcommand{\ee}{\end{equation}}
\newcommand{\barr}{\begin{eqnarray}}
\newcommand{\earr}{\end{eqnarray}}
\newcommand{\breakeq}{\nonumber \\ &&}

\begin{document} 
\draft
\title{Circularly polarized light emission in scanning tunneling
microscopy of magnetic systems} 
\author{S. P. Apell} 
\address{Department of Applied Physics,  Chalmers University of Technology 
          and G\"{o}teborg University,  S-412\,96 G\"{o}teborg, Sweden} 
\author{D. R. Penn}
\address{Electron Physics Group,  National Institute of Standards and
          Technology,  Gaithersburg, MD 20899, USA} 
\author{P. Johansson}
\address{Department of Theoretical Physics,  University of Lund,  
          S\"olvegatan 14 A, S-223\,62 Lund, Sweden}

\date{\today}
\maketitle
\begin{abstract}
Light is produced when a scanning tunneling microscope is used to
probe a metal surface. Recent experiments on cobalt utilizing a
tungsten tip found that the light is circularly polarized;
the sense of circular
polarization depends on the direction of the sample magnetization,
and the degree of polarization is of order 10 \%. This raises the
possibility of constructing a magnetic microscope with very good spatial
resolution. We present a theory of this effect for iron and cobalt and
find a degree of polarization of order 0.1 \%. This is in disagreement
with the experiments on cobalt as well as previous theoretical work
which found order of magnitude agreement with the experimental
results. However, a recent experiment on iron showed 0.0 ${\pm }$2 \%.
We predict that the use of a silver tip would 
increase the degree of circular polarization for a range of photon
energies.
\end{abstract}


\begin{multicols}{2}

\section{Introduction}
\label{SecIntro}
The last ten years have seen a rapid development of the field of
magnetic dichroism,
especially where the response of a system to left and right circularly
polarized light is probed; so called circular dichroism.\cite{Dubs}
For a magnetic material this is magnetic circular
dichroism (MCD). Experiments involve x-ray absorption,\cite{Hague} as well as
standard photo-emission techniques.\cite{Bansmann}

\begin{figure}
\centerline{
\psfig{file=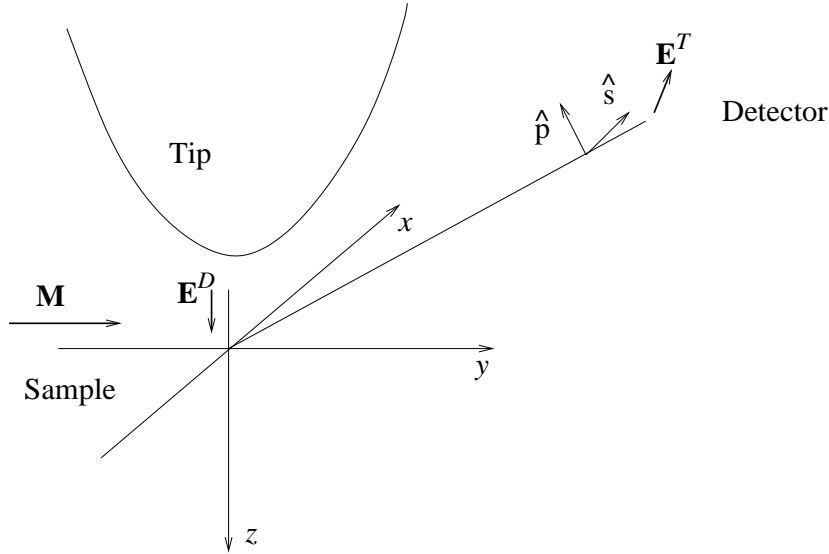,width=4.3in,angle=-90,%
bbllx=158 pt,bblly=178 pt,bburx=453 pt,bbury=614 pt}}
\vspace{1 cm}
\caption{
The schematic experimental set-up in the experiments of V\'{a}zques de
Parga and Alvarado \cite{Vazques} and Pierce et al. \cite{Pierce} 
showing the relative orientation of the
applied magnetic field ({\bf M}) in the plane of the Co(0001)/Fe(001) 
surface (x,y plane), tungsten tip orientation, and optical detection axis.
With the help of the so called reciprocity theorem 
[Eq.\ (\ref{reciproc1eq})]
one can relate the field intensity at the detector due to a current
at the tip (${\bf E}^{T}$(detector)) to the ``detector-generated'' field
intensity between tip and sample (${\bf E}^{D}$). The latter is 
easier to
construct and hence makes it rather straightforward to find the field
we are most interested in; the one in the tip region. $\hat{\bf s}$ and
$\hat{\bf p}$ denote
two orthogonal polarization directions at the detector.
}
\label{FIG1}
\end{figure}

The possibility of a new method of microscopic measurements of surface
magnetism has been suggested by a recent experiment.\cite{Vazques} In
this experiment, circularly polarized 
light emitted from a scanning tunneling microscope (STM) was observed when
the surface of a ferromagnetic material (Co) was probed with a W tip in a
longitudinal configuration (see Fig.\ \ref{FIG1}, the applied magnetic
field is parallel to both
the surface plane and to the plane of light detection). The handedness
of the circular polarization was found to depend on the direction of
the applied field and the degree of polarization was between
5 and 10\%. The results of V\'{a}zques de Parga and Alvarado\cite{Vazques}
looks, at first sight, to have been corroborated by the theoretical work of 
Majlis et al.\cite{Majlis}  In principle such
an effect, if confirmed, should make it possible to map the magnetic
microstructure of a surface by measuring the circular polarization of
the emitted light while scanning the surface because a
STM tip provides very good spatial
resolution.\cite{Resolvnote}
However, a more recent experiment of this type, 
by Pierce et al.\ \cite{Pierce} carried out
on Fe with a W tip, found 0.0 ${\pm }$2 \% circular polarization.

In this paper we calculate the degree of circular polarization for Fe
and Co. We find two contributions to the circular polarization. The
first is due to the Kerr rotation of the light emitted in the tunneling
process. The second contribution is due to the polarization of the
scanning tip by the electric field of the emitted light.
The polarized tip radiates and the radiation undergoes a Kerr rotation.
This second contribution depends on the polarizability of the tip 
as well as the dielectric properties of the sample and
can be significantly larger than the first contribution.

We obtain results which are between one and two orders of
magnitude smaller than that measured by V\'{a}zques de Parga and
Alvarado\cite{Vazques} but consistent with the measurements of Pierce
et al.\cite{Pierce}. Both workers used W tips, but we find that for a
Ag tip, Co and Fe produce a larger degree of circular polarization
for a range of photon energies.

There are other sources of circular polarization that are not magnetic
in origin. For example, experiments by  V\'{a}zques de Parga and
Alvarado\cite{VazquesAlvarado} and theory by Anisimovas and Johansson \cite{Anisimovas}
show that an asymmetric STM tip can produce a degree of
circular polarization on the order of 10 \%.

In section \ref{SecTheory} we develop a theory for the degree of circular
polarization produced by an STM tip in the presence of a magnetic sample
based on the magneto-optic Kerr effect. In order to understand the 
physics of the magnetic circular dichroism in these experiments the
tip is modeled by a dipole. The purpose of the dipole model is solely
to help to understand the physics and is introduced mainly 
for pedagogical reasons. 
In section \ref{SecImp} the theoretical 
description of the tip is improved
for the purpose of obtaining reliable numerical
estimates of the degree of circular polarization. These 
results are discussed in section \ref{SecNumres}.

\section{Theory}
\label{SecTheory}

In this section we obtain an expression for the circular polarization
of light emitted when an STM tip scans a magnetic material. The
calculation is divided into five parts: (A) We first describe the
experiment and express the results in terms of Stokes parameters. (B)
The electromagnetic fields in the tip region are related to the fields
at the detector. (C) A model for the tip is introduced which makes the
problem tractable. (D) The local field outside the tip is related to the
tip polarization, in the presence of a magnetic substrate.
(E) The different contributions are assembled 
and we obtain a theoretical expression for the changes in the circular
polarization due to a change in the direction of the applied magnetic
field. 

\subsection{Theoretical expression for MCD }
\label{SecIIA}
In this section we relate the MCD to the field amplitudes at the detector.
The experiments by V\'{a}zques de Parga and Alvarado \cite{Vazques}
and Pierce et al.\  \cite{Pierce} used the
longitudinal configuration, i.e. the applied magnetic field is
parallel to the plane of incidence and in the surface plane of
Co(0001) [thin film grown on Au(111)] and Fe(001) (whisker), respectively.
Light was detected at an angle of 30$^{\rm o}$ measured from the surface
and tungsten tips were used in both experiments.
The emitted radiation showed circular polarization which changed when
the applied magnetic field was reversed.
Whereas V\'{a}zques de Parga and Alvarado \cite{Vazques}
used a fixed quarter wave plate and carry out the analysis by
a linear polarizer, Pierce et al.  \cite{Pierce} use a rotating quarter wave
plate and a fixed linear analyzer. In both experiments control measurements
were performed
on clean Au(111) samples and yielded no change of the polarization
of the emitted light upon {\em reversal} of the
external magnetic field. The background due to geometric details of
tip-sample junctions such as those discussed in Refs.\ 
\onlinecite{VazquesAlvarado} and \onlinecite{Anisimovas}
and any residual dichroism
of the view port of the UHV system were removed by reversing the
magnetization of the sample.

The results of the two groups were very
different. Pierce et al.  \cite{Pierce} found no magnetization dependent
circular polarization within an experimental uncertainty of
$\pm$2 \% whereas V\'{a}zques de Parga and Alvarado \cite{Vazques} found
values of the order 5-10 \%.  An earlier experiment
using a Ni tip, thus injecting spin-polarized electrons, and a 
Ni polycrystal sample also showed a large MCD upon
reversal of the magnetization of the tip.\cite{Alvarado}

Expressing the degree of polarization $\rho^{\pm}$ 
of the emitted light in terms of Stoke's parameters we have 
\begin{equation}
 \rho^{+} - \rho^{-} = \frac{S^{+}_{3}-S^{-}_{3}}{S_{0}}.
\label{stokeseq}
\end{equation}
The  superscripts on S$_{3}$ and $\rho$ indicate whether the magnetization
is parallel ($+$) or antiparallel ($-$) to the surface projection 
${\bf k}$, of the photon wave vector.
Following the definition given by Jackson \cite{Jackson} and concentrating on 
$\rho^{+}$, left-circular polarized light (positive helicity)
has the polarization vector $\hat{\epsilon}_+=(\hat{p}-i\hat{s})/\sqrt{2}$,
whereas for right-circular polarization
$\hat{\epsilon}_-=(\hat{p}+i\hat{s})/\sqrt{2}$, where, see Fig.\ \ref{FIG1},
$\hat{s}=\hat{x}$ and $\hat{p}=\hat{\theta}$ are s- and p-polarization
unit vectors.
In terms of the electric field ${\bf E}^{\rm T}_{s,p}$ at the detector
caused by the source at the tip (T), the Stokes parameters are 
$$
 S_0\equiv 
      |\hat{\epsilon}_{+}^*\cdot{\bf E}^{\rm T}_{s,p}|^2
    +
      |\hat{\epsilon}_{-}^*\cdot{\bf E}^{\rm T}_{s,p}|^2
   =
    |E_p^{\rm T}|^2
 +
    |E_s^{\rm T}|^2,
$$
and 
$$
 S_3^{+} \equiv
      |\hat{\epsilon}_{+}^*\cdot{\bf E}^{\rm T}_{s,p}|^2
    -
      |\hat{\epsilon}_{-}^*\cdot{\bf E}^{\rm T}_{s,p}|^2
   =
    -2\, {\rm Im } [E_s^{\rm T} E_p^{*{\rm T}}].
$$
As a result, 
\be 
 \rho^{+} = 
    -2\, {\rm Im } \left[
            \frac{ E_s^{\rm T} E_p^{*{\rm T}} }
                 { |E_p^{\rm T}|^2 + |E_s^{\rm T}|^2 }
      \right] 
      \approx
    -2\, {\rm Im } \left[
            \frac{ E_s^{\rm T} }
                 { E_p^{\rm T} }
      \right] ,
\ee
where the last approximation follows because $|E_s^{\rm T}|$ is normally
much smaller than $|E_p^{\rm T}|$ for the particular set-up we are considering.
Hence, the experimentally measured quantity is
\be 
 \rho^{+} - \rho^{-}  
      \approx
    2\, {\rm Im } \left[
            - \, \frac{ E_s^{\rm T}({\rm detector}) }
                 { E_p^{\rm T}({\rm detector}) }
      \right]  - \left( {\bf M} \rightarrow -{\bf M} \right).
 \label{rhodiffeq}
\ee 
The basic physics behind these equations is that
tunneling electrons undergoing inelastic events via
spontaneous emission produce a radiating field
in the vicinity of the tip.
This leads primarily to emission of p-polarized light,
but when reflected in the surface of the magnetic sample it also
gives rise to a small field component in a direction parallel to the
surface due to the Kerr effect. 
We have thus expressed the measured MCD
in terms of the field amplitudes at the detector. 
To facilitate the ensuing calculations, 
we will next relate these tip-induced fields
in the detector region to complementary fields generated by
sources at the detector.

\subsection{Reciprocity theorem}
\label{SecIIB}

In this subsection we use the reciprocity theorem to reformulate 
Eq.\ (\ref{rhodiffeq}).
The electrons tunneling inelastically between tip and sample can emit
photons and are a source of electromagnetic radiation. This radiation
couples to tip and sample and is finally detected far from the tip. We
have previously found \cite{Johansson} that it is convenient
to use the reciprocity theorem of classical electrodynamics
\cite{Kong} in such a situation because it allows the radiated field
to be approximately determined by a nonretarded calculation if
the wavelength of the emitted light is large compared to the relevant tip
extension.
This theorem essentially states that the result of a measurement
is unchanged if the source and field points are interchanged.
Here, the reciprocity theorem can be written as
\begin{equation}
 \sum_{j} \int dV E^{T}_{j}({\bf x}) \ J^{D}_{j}({\bf x}) =
 \sum_{j} \int dV E^{D}_{j}({\bf x}) \ J^{T}_{j}({\bf x}),
\label{reciproc1eq}
\end{equation}
where j denotes the components x, y, and z. The current ${\bf J}^{T(D)}$
at the tip (detector)
is the source for the electromagnetic field ${\bf E}^{T(D)}$. 

Equation (\ref{reciproc1eq}) is valid for media with time-reversal symmetry.
However, a magnetic material 
does not fulfill this condition and one has to use a modified reciprocity
theorem \cite{Kong} whereby Eq.\ (\ref{reciproc1eq}) 
can still be used provided the
true medium is replaced by its complementary medium 
(one with reversed magnetic field). 
If the dielectric
tensor of the real medium is $\epsilon_{ij}$, then the complementary
medium has a dielectric tensor $\epsilon^{c}_{ij}$ which is the
transpose of $\epsilon_{ij}$; $\epsilon^{c}_{ij}=\epsilon_{ji}$. 
The dielectric matrix has the form
\begin{equation}
\epsilon_{ij} = \left( \begin{array}{ccc} \epsilon_S(\omega) &
\epsilon_{1}\cos\phi & -\epsilon_{1}\sin\gamma \sin\phi \\
-\epsilon_{1}\cos\phi & \epsilon_S(\omega) & \epsilon_{1}\cos\gamma
\sin\phi \\ \epsilon_{1}\sin\gamma \sin\phi & -\epsilon_{1}\cos\gamma
\sin\phi & \epsilon_S(\omega)
\end{array} \right) 
\label{epstenseq}
\end{equation}
where the notation $\epsilon_{1}(\omega) \equiv iQ\epsilon_S(\omega)$ is
sometimes used with Q being the so called magneto-optical constant and
$\epsilon_S(\omega)$ is the substrate dielectric function.
The angles $\phi$ and $\gamma$
specify the direction of the applied magnetic field with respect 
to the  surface normal and the plane of incidence. 
We see that the complementary medium corresponds to changing 
the sign of the off-diagonal components. 

To a high degree of accuracy, the current
between the tip and sample is spatially well-localized and
perpendicular to the substrate. Thus we write 
\begin{equation}
 J^{T}_{j}({\bf x}) = \hat{\bf z} J_{o}({\bf \rho},z)
\label{jTeq}
\end{equation}
where $\hat{\bf z}$ is normal to the surface, pointing inwards.
The relative independence
of $E_{z}^{D}$ on position in the surface-tip region has been 
verified by numerical calculation \cite{Berndt} and is also a
key feature of the models used in this paper.
We can take out an average value of {\bf E}$^{D}$ in the
region between tip and sample and write the
right hand side of Eq.\ (\ref{reciproc1eq}) as
\begin{equation}
 E_{z}^{D} \int dV J_{o}(\rho,z) \equiv j_{o} E_{z}^{D}.
\label{reciproc2eq}
\end{equation}
$E_{z}^{D}$ is the perpendicular component of {\bf 
E}$^{D}$, with
respect to the surface plane, i.e. in the z-direction.
The electronic current from the STM flows primarily in the normal
direction with respect to the surface, thus only a perpendicular
component of a ``detector-generated'' field can couple to that tunneling
current. 

On the left hand side of Eq.\ (\ref{reciproc1eq}), 
we insert a current source with two components,
\begin{equation}
 {\bf J}^{D}({\bf x}) = \hat{\bf n} j_{n} \delta({\bf x}-{\bf x}_{D}),
\label{jDeq}
\end{equation}
where {\bf x}$_{D}$ is the detector position and $\hat{\bf n}$ corresponds
to either the direction $\hat{\bf s}$ or $\hat{\bf p}$. 
Notice that these currents
generate two different $E_{z}^{D}$, which we write as
$E_{z}^{n}$ in what follows (n=s or p). 
Eqs.\ (\ref{reciproc1eq}), (\ref{reciproc2eq}) and (\ref{jDeq}) yield
\begin{equation}
 E_{n}^{T}({\rm detector}) j_{n} = j_{o} E_{z}^{n}.
\label{ETdetect}
\end{equation}
Combining Eq.\  (\ref{ETdetect}) with  Eq.\ (\ref{rhodiffeq}) 
and also including the complementary-medium sign  change
we obtain
\begin{equation}
 \rho^{+} \simeq 
  2\, {\rm Im}\left[
   \frac{E_{z}^{s}/j_{s}}{E_{z}^{p}/j_{p}}
                                         \right]= 
  2\, {\rm Im} 
 \left[
  \frac{E_{z}^{s}/E_{s}^{inc}}{E_{z}^{p}/E_{p}^{inc}}
  \right]
\label{rhopluseq}
\end{equation}
and $\rho^{-}$ is  $\rho^{+}(M \rightarrow -M)$.
In the last line we have replaced $j_{s,p}$ by the corresponding
incoming field strengths $E^{inc}_{s,p}$ since s-polarization and
p-polarization represent two orthogonal polarization states. 
Thus the reciprocity theorem makes it possible to express the tip-generated
field at the detector position ($E_{n}^{T}({\rm detector})$) as the
fields in the tip region generated by incoming s- and p-polarized waves.

\subsection{Determination of fields at STM tip}
\label{SecIIC}

In this section we model the STM tip in such a way as to include
the main physical effects and to allow the development of
a formalism for determining the fields at the tip. 
The system of tip and sample is a difficult one to treat for
several reasons. Even if the tip was perfect, in the sense
of having a well-characterized geometrical shape, 
the resulting electromagnetic field problem
has relatively low symmetry. Consequently, we model the tip as a 
polarizable sphere of radius $R$, with a scalar polarizability 
$\alpha_{o}(\omega)$, situated a distance $d = R+D$ from
a surface. Then we replace the sphere with 
a dipole at $d$. In this way we include the relevant physics, such as a
relatively constant field in the region between tip and sample, while
making the problem tractable. Since the circular polarization is a
ratio between two quantities, we hope that a simpler model can capture the
main features of the emitted light.
A dipole model was also used to study the polar Kerr
effect by Kosobukin \cite{Kosobukin} in the context of near field
optics.
In section \ref{SecImp} we improve upon
this by using
the full sphere in the calculations. We
consider the non-magnetic situation in this section and introduce the
magnetic substrate in Sec.\ \ref{SecIID}.

Consider a sphere centered at {\bf d}=(0,0,-d) where -d is the
position of the sphere
outside a
metal surface whose optical reflection can be described 
in terms of its Fresnel reflection coefficients
$\rho_{s}$ and $\rho_{p}$ for s-polarized and p-polarized light,
respectively. The tip (sphere) above the surface is 
{\em replaced} by a {\em point} polarizable dipole with
polarizability $\alpha_{o}(\omega)$. The total electromagnetic field
at the dipole position {\bf E}({\bf d},$\omega$) can be divided into 
two parts {\bf E}$^{\rm ext}$
and {\bf E}$^{t}$: {\bf E}$^{\rm ext}$ is the solution to Maxwell equations
with an incident electromagnetic field and no tip present while {\bf
E}$^{t}$ is a solution when we have no incoming electromagnetic field
but the induced field at the tip plays the role of a {\em source}
term. Assume the point dipole has an
induced dipole moment {\bf P}=({\bf P}$_{\parallel}$,$P_{\perp})$. The 
solution to the {\bf E}$^{t}$ problem can be simplified if
we decompose the induced field at the tip in Fourier components
parallel to the surface ({\bf k}) and note that they play the role of
incoming electromagnetic fields analogous to the situation in the {\bf
E}$^{\rm ext}$-problem, (however, we have to sum over all possible parallel
wave vector components to get the total field). The
(near) field from a dipole can be Fourier
decomposed according to
\begin{equation}
 {\bf E}({\bf x},\omega) = \int \frac{d^{2}k}{(2\pi)^{2}} {\bf
E}({\bf k},z,\omega) e^{i{\bf k} \cdot\bf x }
\label{EFouriereq}
\end{equation}
where {\bf E}({\bf k},z,$\omega$) is the analogue to an
"incoming" electromagnetic field. 
In our case the lowest order magnetic component
is smaller than the electric field by a factor $\omega$d/c $\ll$ 1. 
The total field from the dipole and its image\cite{Apell} is 
expressed in components parallel and perpendicular to the surface as
\begin{equation}
 {\bf E}_{\parallel}^{t}({\bf k},z,\omega) = ip[{\bf E}^{o}_{\parallel}
e^{ipz} +(\rho_{s} \hat{{\bf T}} {\bf E}^{o}_{\parallel} -
\rho_{p}({\bf 1}-\hat{{\bf T}}) {\bf E}^{o}_{\parallel})
e^{-ipz}]
\label{Epareq}
\end{equation}
and
\begin{equation}
E_{\perp}^{t}({\bf k},z,\omega) = -i{\bf k} \cdot {\bf E}^{o}_
{\parallel} (e^{ipz} + \rho_{p} e^{-ipz})
\label{Eperpeq}
\end{equation}
where $\rho_{s}$ and $\rho_{p}$ are the reflection coefficients for s-
and p-polarized light scattered from the surface.. 
In Eqs.\ (\ref{Epareq}) and (\ref{Eperpeq}) we
have introduced
\begin{equation}
{\bf E}^{o}_{\parallel} = \frac{2\pi}{p} e^{ipd}[-{\bf k}P_{\perp} + p
{\bf P}_{\parallel} + \frac{k^{2}}{p} \hat{{\bf T}} {\bf P}_{\parallel}]
\label{Epar0eq}
\end{equation}
and the transverse projection operator in the surface plane is $\hat{{\bf
T}} = {\bf 1} - \hat{\bf k}\hat{\bf k}$, where $\hat{\bf k}$ is a unit vector
along {\bf k}. Furthermore $p^{2}+k^{2}=q^{2}=\omega^{2}/c^{2}$ for
the wave vector {\bf q} = ({\bf k},p) of the incoming field.  In order
to use the reciprocity theorem above we expose the surface and the tip
to an incoming electromagnetic field {\bf E}$^{{\rm inc}}$. In the absence
of the tip the total field would be {\bf E}$^{\rm ext}
={\bf E}^{{\rm inc}}+{\bf E}^{{\rm refl}}$, 
where ${\bf E}^{{\rm refl}}$ is the reflected field.
Upon introducing the dipole, it will develop an induced polarization
\begin{equation}
{\bf P}({\bf d}) = \alpha_{o}(\omega)[{\bf E}^{\rm ext}({\bf d}) + {\bf
E}^{t}({\bf d})]
\label{Pindeq}
\end{equation}
where {\bf d} =(0,0,-d) is the position of the dipole and {\bf
E}$^{t}$({\bf d}) is the image field of the polarized tip due to the
presence of {\bf P}. One can show that \cite{Apell} (see also
Appendix \ref{AppendA}):
\begin{equation}
 {\bf E}^{t}({\bf d}) = F_{\parallel}{\bf P}_{\parallel}
 + F_{\perp}P_{\perp}\hat{{\bf z}}
\label{Eloceq}
\end{equation}
for {\bf P} in the surface plane ({\bf P}$_{\parallel}$) or
perpendicular to it ($P_{\perp}$). We have defined the feed-back, or
image functions,
\begin{equation}
F_{\parallel} = \frac{1}{2} \int_{0}^{\infty} dk k
\frac{e^{2ipd}}{-ip} [q^{2}\rho_{s}(k,\omega) -
p^{2}\rho_{p}(k,\omega)]
\label{Fpareq}
\end{equation}
and
\begin{equation}
F_{\perp} = \int_{0}^{\infty} dk k^{3} \frac{e^{2ipd}}{-ip}
[\rho_{p}(k,\omega)]
\label{Fperpeq}
\end{equation}
after performing an angular integration in the surface plane.
The non-retarded limit for F${_\parallel}$ and F${_\perp}$ is obtained
by letting $c\rightarrow \infty$ with the result that $p$ 
is replaced by $ik$.

Equation (\ref{Pindeq}) is a self-consistency condition on the 
induced
dipole moment representing the tip. Solving for {\bf P} 
we obtain
\begin{equation}
 P_{i} = \frac{\alpha_{o}(\omega)}{1-\alpha_{o}(\omega)F_{i}}
E^{\rm ext}_{i}({\bf d}),
\label{Pisolveq}
\end{equation}
where i=$\parallel$ and $\perp$. The
denominator in Eq.\ (\ref{Pisolveq}) can be included with the field to 
form an effective field acting on the unperturbed tip (dipole) or it can be
included with the bare polarizability $\alpha_{o}$ to form an
effective polarizability.  For
special frequencies, the coupled system exhibits resonances when
Re($\alpha_{o}$F$_{i}$)=1 and $P_{i}$ can be very
large. An explicit demonstration of this is found  \cite{Takemori}
for a sphere outside a surface. 
Combining Eqs. (\ref{Pindeq}) and (\ref{Eloceq}) we obtain the 
total field at the dipole position, 
\begin{eqnarray}
{\bf E}^{tot}({\bf d})
&=& {\bf P}({\bf d})/ \alpha_{o} \nonumber   \\  
&=& {\bf E}^{\rm ext}({\bf 
d}) + {\bf E}^{t}({\bf d}) \nonumber\\
&\equiv&
{\bf E}^{\rm ext}({\bf d})+G_{\parallel}{\bf E}_{\parallel}^{\rm ext}
({\bf d})
+G_{\perp} \hat{z} {\rm E}_{\perp}^{\rm ext}({\bf d}),
\label{Etoteq}
\end{eqnarray}
where $G_{\parallel,\perp}{\bf E}^{\rm ext}$ is the field at the tip, 
due to the image of the tip produced by $P_{\parallel,\perp}$.
We have introduced an image factor 
$G_{i}$ defined as 
\begin{equation}
G_{\parallel,\perp} =
\frac{\alpha_{o}(\omega)F_{\parallel,\perp}}{1-\alpha_{o}(\omega)
F_{\parallel,\perp}}
\label{Gpareq}
\end{equation}
For a substrate characterized by a frequency dependent dielectric
function, $\epsilon_S(\omega)$, we find in the non-retarded limit:
\begin{equation}
 F_{\perp} = 2 F_{\parallel} = \frac{1}{4d^{3}}
\frac{\epsilon_S(\omega)-1}{\epsilon_S(\omega)+1}.
\label{F2Feq}
\end{equation}
In Eq.\ (\ref{Gpareq}), 
$\alpha_{o}$ contains information about the dipole resonances
and F$_{\parallel,\perp}$ contains information about the surface
(sample) resonances ($\epsilon_S$+1=0 corresponds to surface plasmons).
From Eq.\ (\ref{Gpareq}) we see how the tip and sample couple to yield new
eigenmodes at the poles of G$_{\parallel,\perp}$, 
and also a possible field enhancement.
In this respect our
model contains all the features of more refined calculations for the
STM configuration.\cite{Berndt} For the dipole (tip), we use the 
polarizability for a sphere of
radius R and dielectric function $\epsilon_{T}(\omega)$:
\begin{equation}
 \alpha_{o}(\omega) = 
      R^{3}\frac{\epsilon_{T}(\omega)-1}{\epsilon_{T}(\omega)+2}.
\label{poltip}
\end{equation}
In this way one can also
include a more accurate dielectric response of the tip, e.g. using
measured values for $\epsilon_{T}$. The factor 2 in the denominator
is actually (-1+1/n) where n is the so called depolarization factor.
Thus we could also mimic different tip shapes by choosing different
values of n.

In the following discussion we retain Eq.\ (\ref{Etoteq}) as a generic
form for the resulting field at the tip position when an incoming field is 
incident 
on the tip and the sample. Notice also that the only boundary condition
matching is done in the absence of the tip. After this the
self-consistency condition for the induced dipole moment 
[Eq.\ (\ref{Pindeq})] adjusts the total field strength appropriately.

\subsection{Local field near tip with magnetic substrate}
\label{SecIID}

In section \ref{SecIIB} we related the fields at the detector to the fields
of the
inverse problem where light is scattered from the tip. In the previous
section we showed how the incoming field is affected by the presence
of a tip outside a non-magnetic solid. Furthermore, we expressed the
fields in terms of the total field outside the substrate in the
absence of the tip: {\bf E}$^{\rm ext}(\bf d)$ in Eq.\ (\ref{Etoteq}).  
Now we will calculate {\bf E}$^{\rm ext}(\bf d)$, and also consider 
the changes in 
the reflected dipole field, in the presence of a {\em magnetic} sample. We
first address the field felt by a tunneling electron that 
undergoes an inelastic event leading to the light emission.

In the previous section we calculated the induced polarization at the 
tip in a self-consistent manner and obtained the field at the tip
position itself. A tunneling electron will not only feel the incoming 
field and the reflected field of the incoming light,
but the tip-induced image field as well as the direct dipole field from
the tip. In the region
between the actual tip and substrate these fields vary only
slightly since the dipole is far away from the surface, R$>>$D. 
We can write the field
acting on a tunneling electron as (i=$\parallel,\perp$):
\begin{equation}
 \rm E_{i} = F_{o} P_{i} =
F_{o} \alpha_{o} (1+G_{i}) (1+{\rho}_{i}) E_{i}^{inc}
\label{Eztseq}
\end{equation}
where all fields represent average values in between
tip and sample.
F$_{o}$ is a the dipole factor which gives the field strength for
a given polarizability {\bf P}. Notice that if we have
no surface present E = $\rm F_{o,direct} \alpha_{o} E^{inc}$, where
$\rm F_{o,direct}$ is the direct part of F$_{o}$. 
We see that (1+G) plays the role of a field enhancing factor 
for the incoming and scattered electromagnetic fields from the surface
$(1+ \rho ) E^{inc}$. $\alpha_{o}$ is a property of the tip,
$\rm G$ depends on both the sample and the tip, and $\rho$ is a property
of the sample.

To determine {\bf P} we first have to calculate the reflected field
from a magnetic surface and such a reflection involves the 
magneto-optic Kerr effect.
For general angles between applied magnetic field, surface plane 
and plane of incidence there is
an excellent treatment by Zak et al.\cite{Zak} which is
very useful when dealing with light reflection from a magnetic solid.
For a general magnetic configuration the proper form of the dielectric
matrix was given in Eq.\ (\ref{epstenseq}).
The
off-diagonal elements of this tensor carry the information about the
Kerr effect. The origin of the non-diagonal
components is a coupling between the spin of the electrons in the
solid and their orbital momentum due to the atomic potentials
(spin-orbit coupling). The first theoretical calculations of this
effect were carried out by Hulme \cite{Hulme} and Argyres.\cite{Argyres} 
Equation (\ref{epstenseq})
assumes that the dielectric tensor is diagonal when
$\epsilon_{1}(\omega)=0$ (this approximation can be relaxed but it does not
influence our final conclusions). 

For the applied magnetic field in the surface plane we
can use Eq.\ (\ref{epstenseq}) for the general dielectric tensor 
with $\phi=\pi/2$ and
one angle ($\gamma$) suffices to specify the direction of {\bf M} with
respect to the plane of incidence. The dielectric matrix in 
Eq.\ (\ref{epstenseq}) then
simplifies to 
\begin{equation}
\epsilon_{ij} = \left( \begin{array}{ccc} \epsilon_S(\omega) & 0 &
-\epsilon_{1}(\omega)\sin\gamma \\ 
0 & \epsilon_S(\omega) & \epsilon_{1}(\omega)\cos\gamma \\ 
\epsilon_{1}(\omega)\sin\gamma & -\epsilon_{1}(\omega)\cos\gamma & 
\epsilon_S(\omega)
\end{array} \right).
\label{epstenssimple}
\end{equation}
We use this result in the Zak matrix multiplication method,
together
with an expansion to first order in $\epsilon_{1}(\omega)$, since the
off-diagonal elements are small compared to  $\epsilon_{S}(\omega)$
($\mid\epsilon_{1}(\omega)/\epsilon_S(\omega)\mid \ll$1). 
However, these matrix elements provide a coupling between
s-polarization and p-polarization leading to a non-zero Kerr
rotation. For s-polarized incident light (along x-direction),
$E_{s}^{inc}$, our analysis gives
the reflection coefficient for s-polarized light to lowest order in 
$\epsilon_{1}(\omega)$, as
\begin{equation}
\rho_{s} = \frac{p-p_{S}}{p+p_{S}} + O(\epsilon_{1}^{2})
\label{rhoseq}
\end{equation}
which is the standard Fresnel result.\cite{Jackson}
p=$\sqrt{q^{2}-k^{2}}$ and for $q>k$, 
$k=q\sin\theta$ with $\theta$ being the angle
of incidence. 
For $k>>q$, $p\rightarrow ik$.  
$p_{S}=\sqrt{q^{2}\epsilon_S(\omega)-k^{2}}$, with
$\epsilon_S(\omega)$ defined above. Finally q=$\omega$/c, where c is the
light velocity. Apart from the direct reflection of s-polarized light
there is a small but crucial conversion from s- to p-polarized light
with a reflection coefficient
\begin{equation}
 \rho_{ps} =\epsilon_{1}(\omega) \frac{kpq}{p_{S}(p+p_{S})(\epsilon_S p+ p_{S})}
 \sin\gamma +  O(\epsilon_{1}^{2}) \equiv \rho_{L} \sin\gamma
\label{rhopseq}
\end{equation}
It is proportional to $\epsilon_{1}(\omega)$ since it comes from the
off-diagonal response. For large $\epsilon_S$, $\rho_{ps} \propto
\epsilon_{1}/\epsilon_S^{2}$. With Zak's field conventions, 
Eq.\ (\ref{rhopseq})
corresponds to a reflected field amplitude, $-\rho_{ps}
E_{s}^{inc}\sin\theta$, in the z-direction.  It is clear from 
Eq.\ (\ref{rhopseq}) 
that a
general direction of the applied magnetic field in the surface plane
corresponds to replacing $\epsilon_{1}(\omega)$ in the strictly longitudinal
configuration ($\gamma = \pi/2$) by $\epsilon_{1}(\omega)\sin\gamma$ for the
s-polarized case.
Changing the direction of ${\bf M}$ ($\gamma \rightarrow
\gamma +\pi$) changes the sign of $\rho_{ps}$ as it should.

Repeating the same analysis as above for p-polarized incoming light we
find, to lowest order in the off-diagonal elements:
\begin{equation}
\rho_{p} = \frac{\epsilon_S p-p_{S}}{\epsilon_S p+p_{S}} +
\rho_{T}\cos\gamma +
O(\epsilon_{1}^{2})
\label{rhopeq}
\end{equation}
with
\begin{equation}
 \rho_{T} \equiv \frac{-2\epsilon_{1}(\omega)pk}{(\epsilon_S(\omega) p 
 + p_{S})^2}.
\label{rhoTeq}
\end{equation}
The corresponding reflection coefficient for p- to s-polarized
conversion is $\rho_{sp} = - \rho_{ps}$.
The first term in Eq.\ (\ref{rhopeq}) 
is the standard Fresnel reflection coefficient
for a non-magnetic
solid.\cite{Jackson} For later use we need the non-retarded limit
of equations (\ref{rhopeq}) and (\ref{rhoTeq}), viz.:
\begin{equation}
 \rho_{p}^{o} = \frac{\epsilon_S(\omega)-1}{\epsilon_S(\omega)+1}+
\rho_{T}^{o}\cos\gamma+O(\epsilon_{1}^{2}) ,
\label{rhop0eq}
\end{equation}
where
\begin{equation}
\rho_{T}^{o} = \frac{2i\epsilon_{1}(\omega)}{(\epsilon_S(\omega)+1)^2}
\label{rhoT0eq}
\end{equation}
The first term in Eq.\ (\ref{rhop0eq}) is the classical image factor for a 
solid with dielectric function $\epsilon_S(\omega)$.

Making use of the formalism of Zak et al.\cite{Zak}, we find that
for an irradiated magnetic surface we can make the
following replacements with respect to the non-magnetic situation:
\begin{equation}
 \rho_{s} \rightarrow \rho_{s} + \rho_{sp} \equiv \rho_{s} - \rho_{L}
 \sin\gamma  
\label{rhosreplace}
\end{equation}

and
\begin{equation}
 \rho_{p} \rightarrow \rho_{p} + \rho_{ps} \equiv \rho_{p} + \rho_{L}
 \sin\gamma + \rho_{T}\cos\gamma  .
\label{rhopreplace}
\end{equation}
$\rho_{L}$ is defined in Eq.\ (\ref{rhopseq}) and  $\rho_{T}$
is defined in Eq.\ (\ref{rhoTeq}).

The change in reflection factors for a magnetic surface
compared to the non-magnetic situation will also affect the
reflected field from the tip.  We are interested in the
near-field and therefore take the non-retarded limit of $\rho_{L}$;
$\rho_{L}^{o}$:
\begin{equation}
\rho_{L}^{o} = \frac{q\epsilon_{1}(\omega)}{2k(\epsilon_S(\omega)+1)}
\label{rhoL0eq}
\end{equation}
to lowest order. Notice that $\rho_{p}$ has a finite value
[$\rho_{p}^{o}$ + $\rho_{T}^{o}\cos\gamma$, cf.\ Eq.\ (\ref{rhop0eq})] 
in this limit and that $\rho_{s}$ vanishes as
O((q/k)$^{2}$). In what follows, we neglect $\rho_{L}^o \propto q/k$ 
compared to
$\rho_{T}^o$ since $k \gg \omega/c$ in the non-retarded limit. 
Repeating the previous treatment for
a magnetic material one
finds [by letting $\rho_{s}$ and $\rho_{p}$ be transformed according to
Eqs. (\ref{rhosreplace}) and (\ref{rhopreplace})] that Eq.\ (\ref{Eloceq}) 
is replaced by (see Appendix \ref{AppendA}):
\begin{equation}
 {\bf E}_{\parallel}^{loc}({\bf d}) = F_{\parallel}{\bf P}_{\parallel}
 - F_{T}P_{\perp}(\hat{\bf z}\crossprod\hat{\bf M})
\label{Eparloceq}
\end{equation}
and
\begin{equation}
 E_{\perp}^{loc}({\bf d}) = F_{\perp}P_{\perp} + F_{T}{\bf P}_{\parallel} 
\cdot (\hat{\bf z} \crossprod\hat{\bf M})
\label{Eperploceq}
 \end{equation}
where $\hat{\bf M}$ is a unit vector in the surface plane in the direction
of {\bf M} and $\hat{\bf z}$ is normal to the metal and directed
towards it. The coupling factor due to the off-diagonal
response of the medium, F$_{T}$, is given by (non-retarded limit):
\begin{equation}
 F_{T} \equiv \frac{1}{2i} \int dk k^{2} \rho_{T}^o(k,\omega) e^{-2kd} =
 \frac{1}{4d^{3}}\frac{\epsilon_{1}(\omega)}{(\epsilon_S(\omega)+1)^2}
\label{FTLeq}
\end{equation}
using $\rho_{T}^{o}$ from Eq.\ (\ref{rhoT0eq}) in the last line. 
The physics behind the
structure of the above equations is the following.  A perpendicular
dipole [Eq.\ (\ref{Eparloceq})] provides an electro-magnetic field 
which is reflected in the
surface and gives an induced electric field and dipole component
parallel to the surface (due to Kerr response, through F$_T$) and 
perpendicular to the magnetization {\bf M}.
A parallel dipole likewise is reflected and provides an induced
perpendicular field and dipole, due to $F_{T}$. If both {\bf M}
and {\bf P$_{\parallel}$} are parallel there is however no such
contribution.

\subsection{Theoretical results}
\label{SecIIE}

We have now developed all the necessary ingredients for calculating
the circular polarization from Eq.\ (\ref{rhopluseq}). 
First consider the case of incident
s-polarized light at an angle $\theta$. A field $E_{s}^{inc}\hat{\bf s}$ 
(along the
x-direction, $\hat{\bf s} = \hat{\bf x}$) sets up a parallel 
polarization of the tip 
\begin{equation}
{\bf P}_{\parallel} = \alpha_{o} [(1+\rho_{s}) E_{s}^{inc}\hat{\bf x} +
F_{\parallel} {\bf P}_{\parallel} - F_{T}P_{\perp}(\hat{\bf z} \crossprod 
\hat{\bf M})] 
\label{Ppareq}
\end{equation}
Eq.\ (\ref{Pindeq}) and Eq.\ (\ref{Eparloceq}) have been used to derive
Eq.\ (\ref{Ppareq}). There is also a perpendicular polarization
\begin{equation}
 P_{\perp} = \alpha_{o} [-\sin\theta \rho_{ps}^{(-M)}\ E_{s}^{inc} + F_{\perp}
P_{\perp} + F_{T}{\bf P}_{\parallel} \cdot (\hat{\bf z} \crossprod\hat{\bf M})].
\label{Pperpeq}
\end{equation}
In the first term we have indicated that the conversion coefficient $\rho_{ps}$
is now, due to the exact form of the reciprocity theorem, 
to be evaluated in a situation
which is the same as the one considered above if we change the sign of M.
The Kerr coupling between the induced dipole moments of the tip is
included through the coupling function $F_{T}$. 
In Eq.\ (\ref{Pperpeq}) the first
term is the z-component Kerr field set up by the incoming
s-polarized light. The second term is the image from the tip-induced
polarization perpendicular to the surface. Finally the third term is
the image from the parallel induced dipole set up by the incoming
s-polarized wave [Eq.\ (\ref{Ppareq})] 
which is converted to a perpendicular component by
the non-diagonal response of the substrate. The latter is described
through $F_{T}$ which is given above. Neglecting $F_{T}$ in
equation (\ref{Ppareq}), substituting the resulting expression 
in equation (\ref{Ppareq}) for
${\bf P}_{\parallel}$ into equation (\ref{Pperpeq}) and using equations 
(\ref{Pindeq}), (\ref{Etoteq}) and (\ref{Gpareq}) 
gives the field experienced by an electron between tip and sample
to lowest order in $F_{T}$ 
[c.f. Eq.\ (\ref{Eztseq})]:
\begin{eqnarray}
E_{z}^{s} = &&
(1+G_{\perp})(\alpha_{o}F_{o})
\breakeq
 \times
[\rho_{L}\sin\theta - 
(1+\rho_{s}) G_{\parallel} K(\omega)] E_{s}^{inc} \sin\gamma
\label{Ezstseq}
\end{eqnarray}
due to the incoming field $E_{s}^{inc}$. We have used 
$\hat{\bf x} \cdot (\hat{\bf z} \crossprod \hat{\bf M}) = - \sin\gamma$ and
defined:
\begin{equation}
 K(\omega) \equiv F_{T}/F_{\parallel} = 
  \frac{2\epsilon_{1}(\omega)}{(\epsilon_{S}^{2}(\omega)-1)}
\label{Req}
\end{equation}
Note that the 
right hand side of equation (\ref{Ezstseq}) is proportional 
to $\sin\gamma$, the
orientation of the magnetization in the surface plane; longitudinal
polarization providing for the maximum field strength.

$E_{z}^{s}$ in equation (\ref{Ezstseq}) is the major part of the 
average field in the z 
direction in the narrow region between
tip and sample, created by the incoming s-polarized field of magnitude 
$E_{s}^{inc}$.
The first factor in Eq.\ (\ref{Ezstseq}) is understood as follows; 
an incoming s-polarized field of magnitude $E_{s}^{inc}$ undergoes 
a Kerr rotation and a field proportional to $\rho_L$ 
is created in the z direction.
The tip acquires a z-component of polarization and radiates.
This radiation is also reflected back to the tip by the surface, thus 
the tip sees its own image giving a contribution, $ G_{\perp}\rho_{L}$.
The total field at the tip is the original field plus the image field. 
Thus, $G_{\perp}$ is an image
(enhancement) factor due to the polarization of the tip in the
$\hat{\bf z}$ direction. 
The quantity $G_{\perp}$ depends on
the dielectric properties of the tip, the metal, the distance between
them, and on the geometry of the tip. 

The second term in equation (\ref{Ezstseq})
is explained as follows; $(1+\rho_{s})E_{s}^{inc}$ is the field at
the tip due to
light that falls directly on the tip plus light that is reflected from
the metal surface. The tip is then polarized in a direction parallel to
the surface and it radiates. This s-polarized radiation 
is reflected in the surface and undergoes a Kerr rotation so that it
develops a z-component. The term
G$_{\parallel} K$ is the analogue of  
$\rho_{L}$ in the first term. Once a field is created in the z direction 
it is enhanced by the factor $(1 + G_{\perp} )$ in front.

Performing the above calculation for an incoming p-polarized wave with
amplitude $E_{p}^{inc}$ and working only to zeroth order in the
off-diagonal dielectric matrix (since both terms in 
Eq.\ (\ref{Ezstseq}) above for
s-polarized light are already of first order in $\epsilon_{1}(\omega)$) we
find:
\begin{equation}
E_{z}^{p} = - (1+G_{\perp})(\alpha_o F_{o})
(1+\rho_{p})\sin\theta E_{p}^{inc}
\label{Ezptseq}
\end{equation}
where $\sin\theta E_{p}^{inc}$ is the field that falls directly on the
tip and $\rho_{p}\sin\theta E_{p}^{inc}$ is the field at the tip that
is reflected from the surface. The total field is enhanced by the same factor
$(1+G_{\perp})$ as discussed above.

With the use of 
Eqs.\ (\ref{rhodiffeq}), (\ref{rhopluseq}), 
(\ref{Ezstseq}), (\ref{Req}) and (\ref{Ezptseq}) this leads
to the following expression for the magnetic circular dichroism, to lowest
order in $\epsilon_{1}(\omega)$: 
\begin{equation} \rho^{+} - \rho^{-} = 4 \sin\gamma \, {\rm Im}
\left[
 - \frac{\rho_{L}}{1+\rho_{p}} 
  + G_{\parallel} \,
   \left( \frac{1+\rho_{s}}{1+\rho_{p}} \right) \,  \frac{K(\omega)}{\sin\theta}
\right],
\label{rhoresulteq}
\end{equation}
Note that the 
light enhancement factor $G_{\perp}$ 
from the perpendicular field component drops completely out of the problem
as does the dipolar factor F$_{o}$ so the ratio does not depend
on where in the junction the light emission takes place. 
In section \ref{SecImp} we will make a more realistic estimate of 
Eq.\ (\ref{rhoresulteq}) 
for the magnitude of the magnetic circular dichroism.
However, let us already here say that the fact that $G_{\perp}$ drops out
of the dipole-model  calculation gives one hint to why  this model, 
as we will see,
yields results for the magnetic circular dichroism 
that are of the right order of magnitude
although the field enhancement (described by $G_{\perp}$)
in the dipolar model may be very different from that calculated in the
improved model.

\section{Improved theoretical model}
\label{SecImp}

Here we outline a calculation of the tip-induced 
MCD signal within a model geometry where the STM tip is represented
by a sphere characterized by a bulk dielectric function 
$\epsilon_{T}(\omega)$. This allows for a much better description of the
tip polarization.  
Most of the calculational details are deferred
to Appendix \ref{AppendImp}. 

We set out to  determine $E_z^s$ in 
Eq.\ (\ref{rhopluseq}), and this is achieved via the following four steps.
(i) The incoming s-polarized wave is reflected by the 
sample surface.  This yields a total field $E^{\rm ext}$
parallel to the surface. (ii) That field in turn drives the model tip so that it
sends out a field that is reflected back and forth between 
the tip and sample. In this step, we need to extract the part of the 
field that the tip sends onto the sample.
(iii) Next, due to the off-diagonal elements of the 
sample dielectric tensor, part of the electric field parallel to, and 
incident on the sample is converted to a field perpendicular to the 
surface as follows from Eq.\ (\ref{rhop0eq}). 
It is only at this stage that the magnetic properties of the sample 
enters the calculation.
(iv) In the last step, we calculate the degree to which the  
converted electric field is enhanced inside the tip-sample cavity.

{\it Step (i)}: 
With an s-polarized wave incident from the right (positive $y$)
with electric field
$E_s^{\rm inc}$, the Fresnel formulae yields a total 
field just outside the sample surface given by
\be 
 {\bf E}^{\rm ext}={\bf E}_{s}^{\rm inc}(1+\rho_s)=
  \hat{x}E_{s}^{\rm inc}\ \frac{2p}{p+p_S}.
\ee
Thus before introducing the model tip into the problem, we have 
an electric field outside the sample that can be described 
(in the non-retarded limit) by the scalar potential
\be
 \phi^{\rm ext}=-xE^{\rm ext},
\label{phiexteq}
\ee
where 
$ E^{\rm ext}=E_{s}^{\rm inc}[2p/(p+p_S)]$.

{\it Step (ii)}: 
Once the model tip  is introduced into the problem
$\phi^{\rm ext}$ alone is no longer a solution of Laplace's equation 
in the region above the sample, instead another contribution
$\phi_{\rm ind}$ has to be added. 
Using the appropriate boundary conditions for the
${\bf E}$ and ${\bf D}$ fields at the sample and tip surfaces 
$\phi_{\rm ind}$ can be determined.
In the following, we only want to keep the part of $\phi_{\rm ind}$ that 
the tip sends onto the sample.  As we will see in Appendix \ref{AppendImp},
this separation can be done by a simple inspection of the solution.

{\it Step (iii)}: 
We proceed to find the field that is reflected from the sample surface due 
to the second term in Eq.\ (\ref{rhop0eq});  
this is the converted field ${\bf E}^{\rm conv}$.  
Equation (\ref{rhop0eq}) defines a surface response function 
$\chi({\bf k},\omega)=\rho_p^0$,
which in terms of incident and reflected electrostatic potentials is
defined as the ratio 
  $[-\phi^{\rm refl}({\bf k},\omega)/
     \phi^{\rm inc}({\bf k},\omega)]$. 
A further analysis shows that within a non-retarded treatment there
is a local relation between
$\hat{x}\cdot {\bf E}_{\rm ind}^{\rm inc}$ and 
$\hat{z}\cdot{\bf E}^{\rm conv}$,
\be 
 E_{z}^{\rm conv}({\bf \rho}) = 
 \frac{2\epsilon_{1}(\omega)}{(\epsilon_S(\omega)+1)^2}
   \left(\hat{x}\cdot{\bf E}_{\rm ind}^{\rm inc}({\bf \rho})\right).
\label{conveq}
\ee
{\it Step (iv)}: 
In this final step, we calculate the enhancement of the converted
field due to the presence of the model tip.
The converted field discussed above can be represented in terms of a scalar
potential $\phi^{\rm c}$. 
Again, with the tip present,
$\phi^{\rm c}$  alone does not solve Laplace's equation;
it must be supplanted by another contribution
$\phi^{\rm c}_{\rm ind}$.
The calculation determining 
$\phi^{\rm c}_{\rm ind}$ is completely analogous to the one 
carried out in Ref.\ \onlinecite{Johansson}, the only difference being
that $\phi^{\rm c}$ is the 
driving ``force'' in the present case.
Having found $\phi^{\rm c}$ and $\phi^{\rm c}_{\rm ind}$, we evaluate
the corresponding electric field on the symmetry axis.
This is the tip-induced contribution to $E_{z}^{s}$ 
appearing in Eq.\ (\ref{rhopluseq}).

\section{Numerical Results and Discussion}
\label{SecNumres}

We now present numerical results for the dipole model of the tip
based on the expression in
Eq.\ (\ref{rhoresulteq}), and for the sphere
model of the tip discussed in Sec.\ \ref{SecImp}.
We use experimental optical data for the dielectric functions of the
tip and the sample. The off diagonal matrix elements
of the sample dielectric function, $\epsilon_{1}(\omega)$, are
obtained from magneto-optic
Kerr effect measurements.
The literature contains many detailed calculations
and measurements of the Kerr effect (no tip present).
The Kerr effect is an optics effect caused by the spin-orbit
interaction that was
discovered in the last century. The spin-orbit interaction is
small in Fe and Co because the orbital momentum in 3d-metals
is small. It is only in this century that a
microscopic theory has emerged.\cite{Hulme,Argyres,Bennet,Cooper}
The Kerr effect has recently been calculated
by a number of groups for a variety of
elements and compounds (see, e.g. Gasche et al.\ and Delin et al.\
\cite{Gasche,Delin}). Similarly, on the experimental side there
have been a number of measurements from those of Krinchik
and Artem'ev \cite{Krinchik}, whose results we use to obtain 
$\epsilon_{1}(\omega)$,
to the recent results of
Weller et al. \cite{Weller}

Equation (\ref{rhoresulteq}), derived for the dipole model,
consists of two terms;
the first corresponds to the direct Kerr rotation by the sample and the
second to the Kerr rotation of the light produced by the radiating
polarization of the STM tip as previously discussed.
We will refer
to the two effects as the substrate Kerr effect and the tip Kerr effect.

In order to use the dipole model to
make an estimate of the tip-induced Kerr effect we
calculate G$_{\parallel}$ from Eqs.\ (\ref{Gpareq}), (\ref{F2Feq}),
and (\ref{poltip}). The distance between the surface and the tip 
(i.e. from the surface to the sphere)
is very small compared to the tip radius so that $d\sim R$ in Eqs.
(\ref{F2Feq}) and (\ref{poltip}) and G$_{\parallel}$ is determined
by using experimental values for the dielectric functions of the tip
and sample. When comparing our results in Eq.\ (\ref{rhoresulteq}) with the
experiments of V\'{a}zques de Parga and Alvarado \cite{Vazques}
and Pierce et al. \cite{Pierce}, we immediately
recognize that in the longitudinal configuration that they use,
$\hat{\bf s}$ and $\hat{\bf M}$ are perpendicular to each other so
that $\sin\gamma$=1.

Considering the improved (sphere) model, 
the quantities entering
Eq.\ (\ref{rhopluseq}) were calculated as outlined in 
Sec.\ \ref{SecImp} and Appendix \ref{AppendImp}.   
As for the model geometry we here 
used a tip-sample distance $D=$5 \AA, and the tip radius $R$ was set to 
300 \AA.

The results of the calculations for the degree of circular polarization,
$\rho^+ - \rho^-$, are presented in Fig.\ \ref{FIGRes}.
Figure \ref{FIGRes} (a) displays the results obtained with a Co sample,
while Fig.\ \ref{FIGRes} (b) shows the results relevant for a Fe sample.
In each panel, the results for the substrate Kerr contribution (one single curve)
and the tip-induced contribution (four curves) to the degree of polarization
are presented separately.
The tip-induced degree of polarization has been calculated for
Ag and W tips, respectively, using either the dipole model or the
sphere model.
The sphere model gives a larger degree of circular polarization than the
dipole model, because it is a better description of the local
electromagnetic interaction between the tip and the sample.
In our calculational schemes, the tip-induced contribution
ultimately results from a
Kerr rotation of the electromagnetic fields incident from
the tip onto the sample.
The strength of the incident field is determined by how strongly the
tip is excited by the incoming $s$ wave as well as by waves reflected
back and forth between the tip and sample.  The sphere model
allows for a more complete treatment of the repeated reflections than the
dipole model, and therefore gives larger values for
$\rho^+ - \rho^-$. The dipole approximation works fairly well because
the circular polarization is given by the ratio of p and s electric
fields.

The effects of the tip-sample interaction is particularly
pronounced for a silver tip.
An isolated sphere has a dipole-plasmon resonance when the dielectric function
$\epsilon_{T}(\omega)= -2$.
For silver, the dielectric function approaches this
value near 3.5 eV.
Then the model tip becomes highly polarizable and
the feedback mechanism, the waves reflected between the tip and the sample,
described above becomes even more effective.
In connection with this the electromagnetic response functions that
enter our calculations undergo large phase shifts so that
$\rho^+ -\rho^-$ changes sign one or several times.\cite{Erratic}
The relatively small magnitude of the results for the MCD is due to the
factor $(\epsilon_S + 1)^2$ in the denominator in Eqs.\ (\ref{rhoT0eq})
and (\ref{conveq}).  Since $\epsilon_S$ for both Co and Fe is rather large
this suppresses the MCD signal. From bulk arguments one could have assumed that
the circular dichroism should be proportional to $\epsilon_{1}/\epsilon_{S}$.
However the presence of the surface changes the magnitude of both the field
going into the solid to be Kerr rotated and the resulting field going 
out again; in both cases with a factor $1/(\epsilon_{S}+1)$.

It is immediately clear that within the model we have considered,
there is no explanation for the large values for the degree of
light polarization found by
V\'{a}zques de Parga and Alvarado.\cite{Vazques}
The calculated results are an order of magnitude or more smaller
than the experimental results found in Ref.\ \onlinecite{Vazques}.
Even though we use a rather simple model for the geometry as compared
with the complicated, and to a certain degree unknown geometry of
a real STM tip, we cannot see how this could make up for the
very large difference between experimental and theoretical results.
Varying the geometry parameters within reasonable limits (within
a non-retarded formulation the results depend only on R/D) can
change the calculated degree of polarization by a factor of 2
at most. This is also the case when using different sets of
optical data for the dielectric functions entering.
On the other hand, our calculated results are consistent with the
experimental results found by Pierce et al.\cite{Pierce} on Fe samples.
Our results are also small
compared to those obtained in the calculations of Majlis et
al., \cite{Majlis} some of the reasons for this are given in a
footnote. \cite{Majlis_footnote}

Our results are not affected by the assumption of a bulk sample
compared to the very thin Co (100{\AA}) film on
Au \cite{Vazques}, provided it can still be described with bulk
dielectric data.
Calculations by Moog et al. \cite{Moog} indicate that there are
no fundamental changes in actual numbers for films ranging in
thickness between 100 {\AA}ngstr\"{o}m and 400, except a slight
enhancement of the Kerr parameters.

\acknowledgements

The research of two of us (S.P.A. and P.J.) is supported by grants
from the Swedish Natural Science Research Council (NFR).
We appreciate comments from
and discussions with Dan Pierce, Angela Davies, Mark Stiles, Robert
Celotta, G. Mukhopadhyay, Anna Delin, Egidijus Anisimovas and J. Zak.

\appendix

\section{}
\label{AppendA}

In this appendix we show how to evaluate the different angular
integrations required in averaging the local dipole field over
a magnetic surface.

We first have a look at the integrations for the non-magnetic case. In
this instance we have three types of integrals appearing \( \int
d^{2}k (\hat{{\bf T}} \cdot {\bf E}_{\parallel}^{o}) \), \(\int d^{2}k
({\bf k} \cdot {\bf E}_{\parallel}^{o}) \) and \( \int d^{2}k {\bf
E}_{\parallel}^{o} \). They can be expressed in terms of the three
integrals \(\int d^{2}k \hat{\bf k} \), \(\int d^{2}k (\hat{\bf k}
\cdot {\bf P}_{\parallel}) \) and \( \int d^{2}k (\hat{\bf k}
(\hat{\bf k} \cdot {\bf P}_{\parallel})) \).  Performing the angular
integration we find that the first integral is identically zero.
Since the second integral is {\bf P}$_{\parallel}$ dotted with the
first integral it also vanishes. This leaves the third integral
(where $\varphi$ is the angular variable):
\begin{equation}
 \int \frac{d\varphi}{2\pi} \hat{\bf k} 
   (\hat{\bf k} \cdot {\bf P}_{\parallel})
 = \frac{1}{2} {\bf P}_{\parallel}.
\label{App1eq}
\end{equation}
This result is obtained by expressing
$\hat{\bf k}$ as $\hat{\bf x}\cos\varphi+\hat{\bf y}\sin\varphi$ and
$\hat{\bf k} \cdot {\bf P}_{\parallel} = P_{x}\cos\varphi+P_{y}\sin\varphi$.

In the magnetic case we have two extra angular factors in the integrand
compared to the non-magnetic case; $sin\gamma$ and
$\cos\gamma$. Expressed in terms of $\hat{\bf k}$ they are
$\hat {\bf k} \cdot \hat {\bf M}$ and    
$\hat{\bf z}\cdot(\hat{\bf M} \crossprod\hat{\bf k})$,
where $\hat{\bf M}$ is a unit vector in the direction of the
applied magnetic field. Repeating the same steps as above we find that
all integrals can be expressed in terms of the following four
integrals: \( \int d^{2}k (\hat{\bf z}\cdot(\hat{\bf M} \crossprod\hat{\bf 
k}) ) \), \( \int d^{2}k\hat{\bf k}
(\hat{\bf z}\cdot(\hat{\bf M} \crossprod\hat{\bf k}) )) \), \( \int d^{2}k
[(\hat{\bf k} \cdot
{\bf P}_{\parallel}) (\hat{\bf z}\cdot(\hat{\bf M} \crossprod\hat{\bf k}))] 
\) and,\( \int d^{2}k
[\hat{\bf k} (\hat{\bf k} \cdot {\bf P}_{\parallel})
 (\hat{\bf z}\cdot(\hat{\bf M} \crossprod\hat{\bf k}))]
\).  The first and last integrals vanishes since they are odd in
$\hat{\bf k}$. The second integral becomes $\frac{1}{2}
(\hat{\bf z}\crossprod\hat{\bf M})$ using Eq.\ (\ref{App1eq}). 
For the third integral we 
obtain: \begin{equation}
  \int \frac{d\varphi}{2\pi} (\hat{\bf k} \cdot {\bf P}_{\parallel})
(\hat{\bf z}\cdot(\hat{\bf M} \crossprod\hat{\bf k}) ) =
 \frac{1}{2} {\bf P}_{\parallel} \cdot (\hat {\bf z} \crossprod\hat{\bf M})
\label{App2eq}
\end{equation}
again expressing the different vectors in $\hat{\bf x}$ and $\hat{\bf y}$
components and using Eq.\ (\ref{App1eq}).  
With the use of these results it is a 
straightforward manipulation to arrive at 
Eqs.\ (\ref{Eparloceq}) and (\ref{Eperploceq}).

\section{}
\label{AppendImp}

This Appendix explains the calculations outlined in Sec. \ref{SecImp}
in more detail.
To carry them out we use bispherical coordinates,\cite{Morse}
furthermore, to facilitate the connection with similar, earlier calculations
by us\cite{Johansson} and others,\cite{Rendell} we also introduce another 
Cartesian coordinate system ($x',y',z'$) in which $x'$ and $z'$ are reversed
compared with $x$ and $z$, see Fig.\ \ref{Bisphfig}.
The bispherical coordinates $(\beta,\alpha,\varphi')$ are defined by
\be
 z'=\frac{a\sinh{\beta}}{\cosh{\beta}-\cos{\alpha}}, \ \ 
 x'+iy'=\frac{a\sin{\alpha}\ e^{i\varphi'}}{\cosh{\beta}-\cos{\alpha}}.
\ee
Both the sample surface ($z'=0$, $\beta=0$) 
and the sphere 
$$\beta=\beta_0=\ln\left({1+(D+a)/R}\right),$$
are constant $\beta$ surfaces. 
The length scale is set by the parameter
$$a=\sqrt{D^2+2RD}.$$

Let us embark on the calculations described in Sec.\ \ref{SecImp}. 
Step (i) is a straightforward application of the Fresnel
formulae that yields an external potential
$\phi^{\rm ext}=-xE^{\rm ext}=x'E^{\rm ext}$ 
[see Eq.\ (\ref{phiexteq})]
describing the electric field of an $s$ polarized wave 
reflected off the sample surface.
In the bispherical coordinates this potential 
can be written 
\barr 
 \phi^{\rm ext}= && -E^{\rm ext}\ 
  \sqrt{8}\ a\ \sqrt{\cosh \beta -\cos \alpha}
\breakeq
\times
 \sum_{1}^{\infty} e^{-(n+1/2)|\beta|} P_{n}^1(\cos \alpha) \cos \varphi'.
\earr
Proceeding to step (ii), we introduce  the induced potential
$\phi_{\rm ind}$, and  make the Ansatz
\barr 
 \phi_{\rm ind}=&& - 
  \sqrt{8}\ a\ E^{\rm ext} \sqrt{\cosh \beta -\cos \alpha}
\breakeq
\times
 \sum_{1}^{\infty} F_n(\beta) P_{n}^1(\cos \alpha) \cos \varphi'.
\label{phiindeq}
\earr
In the sample ($\beta\le 0$), the function $F_n(\beta)$ is given by
\begin{mathletters}
\be
 F_n(\beta)=(A_n+B_n)e^{(n+1/2)\beta},
 \label{Fneqsamp}
\ee
in the tip ($\beta\ge\beta_0$)
\be
 F_n(\beta)=(A_ne^{(2n+1)\beta_0}+B_n)e^{-(n+1/2)\beta},
 \label{Fneqtip}
\ee
and finally, {\em between} the sample and tip ($0 \le \beta \le \beta_0 $)
\be 
 F_n(\beta)=A_ne^{(n+1/2)\beta} + B_n e^{-(n+1/2)\beta}.
 \label{Fneq}
\ee
\end{mathletters}\noindent
Thus, $F_n$, $\phi_{\rm ind}$, and therefore the tangential 
${\bf E}$ field is continuous across the tip and sample interfaces.
From the form of Eq.\ (\ref{Fneq}) it is clear that the field that
the tip sends onto the sample is contained in the $A_n$ terms;
these fields decay exponentially as one goes away from the tip.

To determine the coefficients, $A_n$ and $B_n$ we must also
demand that the displacement field perpendicular to
the sample and tip surfaces is continuous across these interfaces.
At the sample this means that
\be 
  \epsilon_S
     \frac{\partial \phi_{\rm ind}}{\partial \beta}|_{\beta=0-}
 =
  \frac{\partial \phi_{\rm ind}}{\partial \beta}|_{\beta=0+}
\ee
(here $\epsilon_S$ is the sample dielectric function),
which yields 
\be
  B_n=-\chi_SA_n,
\ee
where the sample surface response function 
\be 
 \chi_S=
  \frac{\epsilon_S - 1}{\epsilon_S+1}.
\label{chisdef}
\ee 
Note that the contribution to
$D_{\perp}$ coming from $\phi^{\rm ext}$ is already continuous
since the external potential results from using the Fresnel formulae.
At the tip-vacuum interface
($\beta=\beta_0$),
both $\phi_{\rm ind }$ and $\phi^{\rm ext}$ must be considered
in the boundary condition for $D_{\perp}$.
This yields an equation system involving the coefficients $A_n$
\be
 U_n^sA_n+V_n^sA_{n-1}+W_n^sA_{n+1}=S_n,
\label{seqsyst}
\ee
where
\begin{mathletters}
\barr
  U_n^s &&=
    -(2n+1) \cosh \beta_0 \, \left(
         e^{(n+1/2)\beta_0} - \chi_S\, \chi_T\, e^{-(n+1/2)\beta_0} \right),
\breakeq
      + \sinh\beta_0 \,\chi_T 
          \left(e^{(n+1/2)\beta_0}- \chi_S\, e^{-(n+1/2)\beta_0}\right)
\earr
\be
 V_n^s=(n-1)
  \left(e^{(n-1/2)\beta_0}- \chi_S\, \chi_T\, e^{-(n-1/2)\beta_0}\right),
\ee
\be
 W_n^s=(n+2)
   \left(e^{(n+3/2)\beta_0}- \chi_S\, \chi_T\, e^{-(n+3/2)\beta_0}\right),
\ee
and
\barr
 S_n&&=-\chi_T \left \{ e^{-(n+1/2)\beta_0}
     \left[ \sinh \beta_0 -(2n+1) \cosh \beta_0 \right] \right.
 \breakeq
  \left.
    +(n-1) e^{-(n-1/2)\beta_0}
    +(n+2) e^{-(n+3/2)\beta_0}
   \right \} .
\earr
\end{mathletters}\noindent
One arrives at these equations through a procedure that 
is completely analogous to
the one used in earlier calculations.\cite{Johansson,Rendell}
Solving Eq.\ (\ref{seqsyst}) for the $A_n$ coefficients, we can determine
the tip-induced potential {\em incident} on the sample,
\barr 
 \phi_{\rm ind}^{\rm inc}=&& - 
  \sqrt{8}\ a\ E^{\rm ext} \sqrt{\cosh \beta -\cos \alpha}
\breakeq
\times
 \sum_{1}^{\infty} A_ne^{(n+1/2)\beta} P_{n}^1(\cos \alpha) \cos \varphi'.
\label{phiindinceq}
\earr

Next, we have to consider the electric field conversion at the sample surface
due to the off-diagonal components of the dielectric tensor 
in Eq.\ (\ref{epstenssimple}).
In the primed coordinate system, it takes the form
\begin{equation}
  \epsilon'_{ij} = \left( \begin{array}{ccc} 
  \epsilon_S(\omega) & 0 & -\epsilon_{1}(\omega)\sin\gamma \\ 
  0 & \epsilon_S(\omega) & -\epsilon_{1}(\omega)\cos\gamma \\ 
  \epsilon_{1}(\omega)\sin\gamma & \epsilon_{1}(\omega)\cos\gamma & 
  \epsilon_S(\omega)
  \end{array} \right). 
\label{epstensrotated}
\end{equation}
Let us use this to calculate the modified sample surface response function.
If a potential
\be
 \phi^{\rm inc}= e^{i{\bf k}\cdot{\bf \rho}} e^{kz'}e^{-i\omega t}
\ee
acts on the sample, its response yields another
contribution to the potential above the surface
\be
 \phi^{\rm refl}= -\chi_S({\bf k},\omega) 
     e^{i{\bf k}\cdot{\bf \rho}} e^{-kz'}e^{-i\omega t},
\ee
where $\chi_S({\bf k}, \omega)$ is the surface response function.
Combining these two contribution to the potential with a solution
to Laplace's equation inside the sample
\be
 \phi^{\rm tr}= T({\bf k},\omega) 
     e^{i{\bf k}\cdot{\bf \rho}} e^{kz'}e^{-i\omega t},
\ee
and applying the usual boundary conditions, we obtain using
Eq.\ (\ref{epstensrotated}) that
\be
 \chi_S(\omega)=\frac{\epsilon_S-1}{\epsilon_S+1} +
              \frac{2i\epsilon_1 \sin (\gamma+\varphi'_k)}{(\epsilon_S+1)^2},
\label{surfresp}
\ee
where $\varphi'_{k}$ is the angle between ${\bf k}$ and $\hat{x'}$.

The second term in Eq.\ (\ref{surfresp}) is  our main concern;
it governs the electric field conversion at the surface of the
magnetic sample.  In view of Eq.\ (\ref{surfresp}) we write  
the potential corresponding to the converted field as 
\be
 \phi^{\rm conv}({\bf k}) =
  - \frac{2i\epsilon_{1}(\omega)}{(\epsilon_S+1)^2}
   \sin (\gamma+\varphi'_k) \, \phi_{\rm ind}^{\rm inc} ({\bf k}).
\ee
Since we do not know the Fourier transform 
$\phi_{\rm ind}^{\rm inc}({\bf k})$ 
we cannot immediately use this relation.
However, specializing to the longitudinal configuration ($\gamma=\pi/2$)
and using the fact that 
on the sample surface, $ \phi_{\rm ind}^{\rm inc}=f(\rho) \cos \varphi'$
($\rho$ is the distance to the symmetry axis),
one can show that with ${\bf M}=M\hat{y'}$ there is a local relation between 
$\hat{x'}\cdot{\bf E}_{\rm ind}^{\rm inc}$ 
and $\hat{z'}\cdot{\bf E}^{\rm conv}$,
\be 
 E_{z'}^{\rm conv}({\bf \rho})= 
 \frac{2\epsilon_{1}(\omega)}{(\epsilon_S+1)^2}
   \left(\hat{x'}\cdot{\bf E}_{\rm ind}^{\rm inc}({\bf \rho})\right).
\label{conveq2}
\ee
The calculation leading to Eq.\ (\ref{conveq2}) also shows that
both these fields can be written on the form
$f_0(\rho) + f_2(\rho) \cos 2\varphi'$, where $f_0$ and $f_2$ are
functions of $\rho$.
The second term  obviously cannot
induce an electric field possessing a 
$z'$ component on the symmetry axis of the 
tip-sample system, so we neglect it from now on.
The cylindrically symmetric part of the  field 
$ {\bf E}^{{\rm conv}}({\bf \rho})$ can be derived from a 
potential written as
\be
 \phi^{\rm c}=\sqrt{2}\, a \sqrt { \cosh \beta -\cos \alpha }
 \sum_{0}^{\infty} T_n e^{-(n+\frac{1}{2})\beta} P_n (\cos \alpha).
\label{phiconvsymm}
\ee

The coefficients $T_n$ in Eq.\ (\ref{phiconvsymm}) can be calculated 
with the aid of Eq.\ (\ref{conveq2}).
To this end we extract the cylindrically symmetric part of
$\hat{x'}\cdot{\bf E}_{\rm ind}^{\rm inc}$ by taking the average of its 
values on the $x'$ and $y'$ axis, respectively, and rewrite these 
expressions as sums of Legendre polynomials only. 
At the expense of introducing a more complicated prefactor this yields
\be
 \hat{x'}\cdot {\bf E}_{\rm ind}^{{\rm inc},x'(y')}=
  \sqrt{8}\  E^{\rm ext} 
 \frac{\sqrt{1-\cos \alpha}}{1+\cos \alpha}
 \sum_{0}^{\infty} C_n^{x'(y')}  P_{n}(\cos \alpha),
\label{eonaxis}
\ee
where the superscript $x'(y')$ indicates whether the field 
is evaluated on the $x'$ or $y'$ axis.
The coefficients $C_n^{x'}$ and $C_n^{y'}$ are given by rather lengthy
expressions involving the $A_n$ coefficients.
It is also possible to rewrite  $E_{z'}^{\rm c}$ resulting
from the $z'$ gradient of Eq.\ (\ref{phiconvsymm}) as a sum 
of Legendre polynomials preceded by the same prefactor as in Eq.\
(\ref{eonaxis}).  From Eq.\ (\ref{conveq2}) 
we then obtain the following equation system determining the $T_n$'s
\be
 v_n T_{n-2} + u_n T_{n} +w_n T_{n+2} = 
 \frac{\epsilon_{1}(\omega)}{(\epsilon_S+1)^2} E^{\rm ext} (C_n^{x'}+ C_n^{y'}),
\ee
where 
\barr
  && v_n = -\frac{1}{4}\frac{(n-1)n}{2n-1},
\breakeq
u_n =\frac{1}{4} \left ( 2n+1-\frac{(n+1)^2}{2n+3} 
                           -\frac{n^2}{2n-1} \right ),
\ \ {\rm and}
\breakeq
w_n = - \frac{1}{4} \frac{(n+1)(n+2) } {2n+3}.
\earr

The final step amounts to solving for the enhancement of the $z'$ component
of the electric field between the tip and sample due to 
the presence of the tip.  
In fact, to obtain the final result for the degree of circular
polarization, this calculation has to be done with two different 
driving forces. On one hand, using $\phi^{\rm c}$
as a driving force we obtain  $E_z^s$ to be used in Eq.\ (\ref{rhopluseq}),
if instead, as in Ref.\ \onlinecite{Johansson}, the potential
corresponding to a reflected $p$ wave is used to drive the tip-sample
system, we obtain $E_z^p$ to be used in Eq.\ (\ref{rhopluseq}).
The two calculations can be done in parallel.
We make an Ansatz for the potentials induced due to 
the presence of the tip in the two cases
\be 
\phi_{\rm ind}^{\rm c(p)}=
 \sqrt{2}\, a \sqrt{\cosh \beta -\cos \alpha }
 \sum_{0}^{\infty} F_n^{\rm c(p)}(\beta) P_n (\cos \alpha),
\ee
where the superscripts c and p, respectively, indicate that it is either
the converted field or an incoming $p$ wave that plays the role of driving 
force.
\begin{mathletters}
In the sample ($\beta \le 0$),
\be 
  F_n^{\rm c(p)}(\beta) = 
 A_n^{\rm c(p)} e^{(n+1/2) (\beta-\beta_0)} 
  + B_n^{\rm c(p)} e^{(n+1/2)(\beta+\beta_0)},
\ee
in vacuum    ($ 0  \le \beta \le \beta_0 $)
\be
 F_n^{\rm c(p)}(\beta) = 
 A_n^{\rm c(p)} e^{(n+1/2) (\beta- \beta_0) }+ 
   B_n^{\rm c(p)} e^{-(n+1/2)(\beta - \beta_0)}, 
\ee
and in the tip ($\beta \ge \beta_0$),
\be
 F_n^{\rm c(p)}(\beta) = \left( A_n^{\rm c(p)} + B_n^{\rm c(p)} \right ) 
                               e^{-(n+1/2)(\beta-\beta_0)}.
\ee
\end{mathletters}\noindent
The external potential in the  case of the converted field is given by
Eq.\ (\ref{phiconvsymm})
$$
 \phi^{\rm c}=\sqrt{2}\, a \sqrt { \cosh \beta -\cos \alpha }
 \sum_{0}^{\infty} T_n e^{-(n+\frac{1}{2})\beta} P_n (\cos \alpha),
$$
whereas in the case of an incoming $p$ wave it is 
\barr 
 \phi_p^{\rm ext}= && - \sin \theta 
             \frac{2\epsilon_S p}{\epsilon_S p+ p_t} E_p^{\rm inc} 
         \sqrt{2} a \sqrt{\cosh \beta-\cos \alpha}
 \breakeq
 \times
           \sum_0^{\infty} (2n+1) e^{-(n+1/2)\beta} P_n(\cos \alpha),
\earr
where the first factor $\propto E_p^{\rm inc}$ is the resulting $z'$
component of the ${\bf E}$ field just outside the sample surface as
obtained from the Fresnel formulae.

Proceeding along the same lines as when deriving Eq.\ (\ref{seqsyst}),
we first obtain
\be 
 B_n^{\rm c(p)}= - A_n^{\rm c(p)} e^{-(2n+1)\beta_0} \chi_S
 \label{An2Bn}
\ee
and then the equation systems
\be
 U_nA_n^{\rm c}+V_nA_{n-1}^{\rm c}+W_nA_{n+1}^{\rm c}=S_n^{\rm c}
\label{ceqsyst}
\ee
and
\be
 U_nA_n^{\rm p}+V_nA_{n-1}^{\rm p}+W_nA_{n+1}^{\rm p}=S_n^{\rm p}
\label{peqsyst}
\ee
determining the enhancement due to the presence of the tip
of the converted field and the $p$ polarized wave, respectively.
The coefficients on the left hand side are the same in both cases,
\begin{mathletters}
\barr
 U_n = &&  (2n+1) \cosh\beta_0 
        \left( 1- \chi_S \chi_T e^{-(2n+1) \beta_0} \right) 
 \breakeq
         -\chi_T \sinh \beta_0 \left(1-\chi_S e^{-(2n+1)\beta_0} \right),
\earr
\be
 V_n= -n \left( 1 - \chi_S \chi_T e^{-(2n-1)\beta_0} \right),
\ee
\be
 W_n=-(n+1) \left(1 - \chi_S \chi_T e^{-(2n+3)\beta_0} \right).
\ee
\end{mathletters}\noindent
The right hand side in Eq.\ (\ref{ceqsyst}) is 
\barr
 S_n^{\rm c}  =&&
  \chi_T\, e^{-(n+1/2) \beta_0} \left\{
    [\sinh\beta_0 -(2n+1)\cosh \beta_0 ] T_n 
   \right.
   \breakeq
   \left.
 +n e^{\beta_0} T_{n-1} + (n+1) e^{-\beta_0} T_{n+1}  
 \right\} ,
\earr
while for the case of an incident $p$ wave we obtain
\barr
 S_n^{\rm p} &&=
  \sin \theta\, \frac{2\epsilon_S p}{\epsilon_S p+ p_t} E_p^{\rm inc} \,
  \chi_T\,  e^{-(n+1/2) \beta_0}
 \breakeq
 \times
 \left \{
  (2n+1)\left[ (2n+1)\cosh \beta_0 - \sinh \beta_0 \right] 
 \right.
 \breakeq
 \left.
  - (2n^2 - n)e^{\beta_0} - (2n^2 + 5n + 3)  e^{-\beta_0}
 \right\}.
\earr

Once all the coefficients $A_n^{\rm c}$ and 
$A_n^{\rm p}$ have been determined using Eqs.\ (\ref{An2Bn}),
(\ref{ceqsyst}), and (\ref{peqsyst}),
we can calculate the resulting electric fields on the symmetry axis
as
\be
 E_{z'}^s=-\frac{\cosh \beta-\cos \alpha}{a} 
    \left[ \frac{\partial \phi^{\rm c}}{\partial \beta}
          + \frac{\partial \phi_{\rm ind}^{\rm c}}{\partial \beta} 
    \right ],
\ee
and
\be
 E_{z'}^p=-\frac{\cosh \beta-\cos \alpha}{a} 
    \left[ \frac{\partial \phi_p^{\rm ext}}{\partial \beta}
          + \frac{\partial \phi_{\rm ind}^{\rm p}}{\partial \beta} 
    \right ].
\ee
The degree of polarization of the emitted light is found from
\be 
 \rho^+= \frac{S_3}{S_0}
      \approx -  2\, {\rm Im}\left[\frac{E_s^{out}}{E_p^{out}} \right],
\ee
so that using the reciprocity theorem, keeping in mind the sign change of 
$\epsilon_1$ discussed earlier 
\be 
 \rho^{+} - \rho^{-} \approx
     4 \, {\rm Im}\left[\frac{E_{z'}^s}{E_{z'}^p} \right].
\ee

\end{multicols}

\vspace{10mm}

\begin{figure}
\input Fig2a.tex
\vspace{1 cm}
\input Fig2b.tex
\vspace{1 cm}
\caption{
Calculated results for the circular polarization for (a) a Co sample,
and (b) a Fe sample.  Each of the figures present results obtained from both
the dipole model (dotted curves) and the improved (sphere) model (full curves),
moreover, results corresponding to Ag and W tips (as  indicated next to the 
curves) as well as the substrate
Kerr contribution are displayed in both panels.
In the sphere-model calculation we used $R=$ 300 \AA\ for
the tip radius and $D=$ 5 \AA\ for the tip-sample separation.
The sphere model in general gives a larger tip-induced contribution
to the degree of polarization, nevertheless the results obtained
with the two different methods have many qualitative features
in common.
}
\label{FIGRes}
\end{figure}

\begin{figure}
\centerline{
\psfig{file=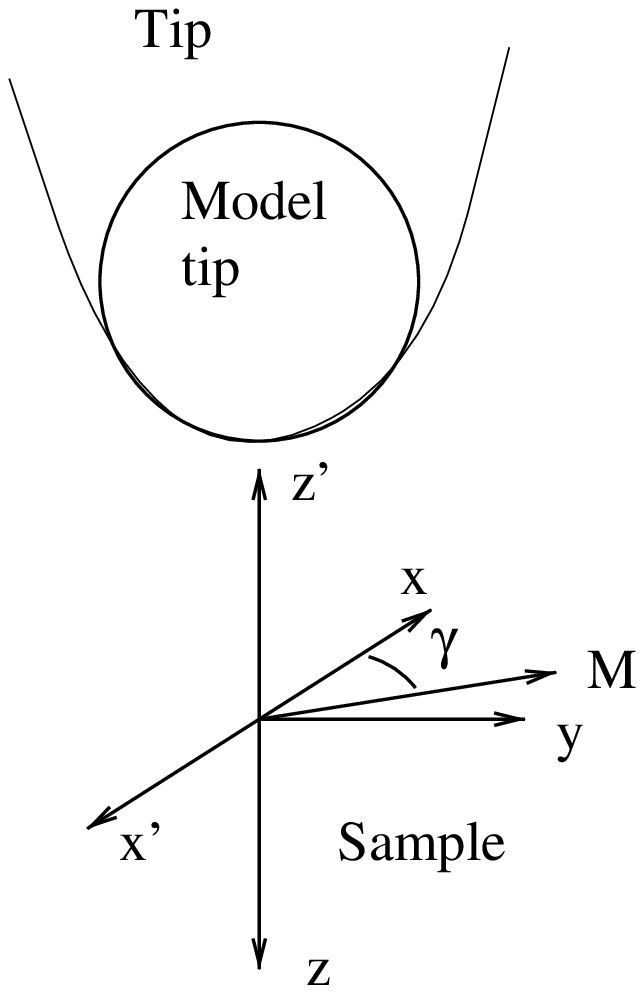,width=2.3in,%
bbllx=215 pt,bblly=254 pt,bburx=397 pt,bbury=538 pt}}
\vspace{1 cm}
\caption{
Schematic illustration of the coordinate system used in the 
improved model calculation.
}
\label{Bisphfig}
\end{figure}


\begin{references}
\bibitem{Dubs}
 R. L. Dubs, S. N. Dixit, and V. McKoy, Phys.\ Rev.\ Lett.\ {\bf 54}, 1249
 (1985).

\bibitem{Hague}
C. F.  Hague, J.-M. Mariot, P. Strange, P. J. Durham, and B. L. Gyorffy,
 Phys.\ Rev.\ B {\bf 48}, 3560 (1993).

\bibitem{Bansmann}
 J. Bansmann, M.  Getzlaff, C. Westphal, F. Fegel, and
 G. Sch\"{o}nhense, Surf.\ Sci.\ {\bf 269/270}, 622 (1992).

\bibitem{Vazques}
 A. L. V\'{a}zques de Parga and S. F. Alvarado, Phys.\ Rev.\ Lett.\ {\bf
 72}, 3726 (1994).

\bibitem{Majlis}
 N. Majlis, A. Levy Yeyati, F. Flores, and R. Monreal,
 Phys.\ Rev.\ B {\bf 52}, 12505 (1995).

\bibitem{Resolvnote} Within the theory we will present here,
the resolution would mainly be determined by the 
size of the electromagnetic cavity formed between the tip and sample.
This leads to a resolution limit of the order of 50--100 \AA.


\bibitem{Pierce}
 D. T. Pierce,  A. Davies, J. A. Stroscio, and R. J. Celotta,
 Appl.\ Phys.\ A {\bf 66}, S403 (1998).

\bibitem{VazquesAlvarado}
A. L. V\'{a}zques de Parga and S. F. Alvarado, Europhys.\ Lett.\ {\bf
36}, 577 (1996).

\bibitem{Anisimovas}
E. Anisimovas and P. Johansson,  Phys.\ Rev.\ B {\bf 59}, 5126 (1999).

\bibitem{Alvarado}
 S. F. Alvarado, in {\em Near Field Optics}, NATO Advanced
 Research Workshop on Near Field Optics, Arc-et-Senans, France, 1992,
 Ser.\ E, Vol.\ 242, edited by D. W. Pohl and D. Courjon (Kluwer,
 Dordrecht, 1993) p.\ 361.

\bibitem{Jackson}
 J. D. Jackson, {\it Classical Electrodynamics} (Wiley,
 New York, 1999).

\bibitem{Johansson}
 P. Johansson, R. Monreal, and P. Apell, Phys.\ Rev.\ B {\bf 42}, 9210
 (1990);
 P. Johansson and R. Monreal, Z. Phys.\  B {\bf 69}, 284 (1991).

\bibitem{Kong}
 J. A. Kong, {\it Theory of Electromagnetic Waves} 
 (Wiley, New York, 1975), section 7.2.

\bibitem{Berndt}
 R. Berndt, J. K. Gimzewski, and P. Johansson, Phys.\ Rev.\ Lett.\ {\bf
 67}, 3796 (1991).

\bibitem{Kosobukin}
V.A. Kosobukin, Physics of the Solid State, {\bf 39}, 488 (1997).  

\bibitem{Apell}
P. Apell, Physica Scripta {\bf 24}, 795, (1981).

\bibitem{Takemori}
 T. Takemori, M. Inoue, and K. Ohtaka, J. Phys.\ Soc.\ Jpn.
 {\bf 56}, 1587 (1987).

\bibitem{Zak}
 J. Zak, E. R. Moog, C. Liu, and S. D. Bader, Phys.\ Rev.\ B {\bf 43},
 6423. Erratum, 1992, Phys.\ Rev.\ B {\bf 46}, 5883 (1991).

\bibitem{Hulme}
 H. Hulme, Proc.\ R. Soc.\ London {\bf 135}, 237 (1932).

\bibitem{Argyres}
 P. N. Argyres,  Phys.\ Rev.\ {\bf 97}, 334 (1955).

\bibitem{Bennet}
 H. S. Bennet and E. A. Stern, Phys.\ Rev.\ {\bf 137}, A448 (1965).

\bibitem{Cooper}
 B. R. Cooper, Phys.\ Rev.\ {\bf 139}, A1504 (1965).

\bibitem{Gasche}
 T. Gasche, M. S. S. Brooks, and B. Johansson, Phys.\ Rev.\ B {\bf 53}, 296
 (1996).

\bibitem{Delin}
 A. Delin, O. Eriksson, B. Johansson, S. Auluck, and J. M. Wills, 1998
(preprint).

\bibitem{Krinchik}
 G. S. Krinchik and V. A. Artem'ev, 
  Sov.\ Phys.\ JETP {\bf 26}, 1080 (1968).

\bibitem{Weller}
 D. Weller, G. R. Harp, R. F. C. Farrow, A. Cebollada, and J. Sticht,
 Phys.\ Rev.\ Lett.\ {\bf 72}, 2097 (1994).

\bibitem{Stearns}
 M. B. Stearns, in 
 Landolt-B\"{o}rnstein New Series III/19a, 
edited by H. P. J. Wijn (Springer-Verlag, Berlin, 1986), pp.\ 113.


\bibitem{Erratic}
The rapid behavior around $\hbar \omega=$ 3.6 eV of the 
calculated results for a Ag tip in the  sphere model is due to
a dramatic decrease in the response to an incoming 
$p$ polarized wave 
(cf.\ Ref.\ \protect\onlinecite{Johansson}),
which makes the degree of polarization extremely sensitive 
even to small variations, for example interpolation errors, 
of the dielectric functions.
We wish to point out that these oscillations in $\rho^+-\rho^-$ 
can definitely not be observed experimentally, for the intensity of 
the emitted light for photon energies above 3.3--3.4 eV is 
extremely small.


\bibitem{Majlis_footnote}
The dielectric tensor of Majlis et al.\cite{Majlis} is unphysical. The
off-diagonal elements are constructed from the diagonal ones, 
an approach for
which there is no theoretical justification, and the constructed
tensor does not obey
proper symmetry conditions for an absorbing medium. Furthermore
the corresponding expression to our $\rho_{ps}$ is not correct.

\bibitem{Moog}
 E. R. Moog, S. D. Bader, and J. Zak, 
  Appl.\ Phys.\ Lett.\ {\bf 56}, 2687 (1990).

 \bibitem{Morse} P. M. Morse and H. Feschbach,
  {\it Methods of Theoretical Physics} 
  (McGraw-Hill, New York, 1953), Vol II, pp. 1298--1301.

 \bibitem{Rendell} R. W. Rendell and D. J. Scalapino,
  Phys.\ Rev.\ B {\bf 24}, 3276 (1981).

\bibitem{Stern}
 E. A. Stern, J. C. Mc Groddy, and W. E. Harte, Phys.\ Rev.\ {\bf 135},
 A1306 (1964).

\end{references}
\end{document}